\documentclass[aps,prl,reprint,superscriptaddress]{revtex4-1}
\usepackage{graphicx}
\usepackage{braket}
\usepackage{braket}
\usepackage{amsmath}
\usepackage[style=english,english=american]{csquotes}
\usepackage{bm}
\usepackage{url}

\begin{document}

\title{Microwave cavity detected spin blockade in a few electron double quantum dot}

\author{A. J. Landig}
\affiliation{Department of Physics, ETH Z\"urich, CH-8093 Z\"urich, Switzerland}
\author{J. V. Koski}
\affiliation{Department of Physics, ETH Z\"urich, CH-8093 Z\"urich, Switzerland}
\author{P. Scarlino}
\affiliation{Department of Physics, ETH Z\"urich, CH-8093 Z\"urich, Switzerland}
\author{C. Reichl}
\affiliation{Department of Physics, ETH Z\"urich, CH-8093 Z\"urich, Switzerland}
\author{W. Wegscheider}
\affiliation{Department of Physics, ETH Z\"urich, CH-8093 Z\"urich, Switzerland} 
\author{A. Wallraff}
\affiliation{Department of Physics, ETH Z\"urich, CH-8093 Z\"urich, Switzerland} 
\author{K. Ensslin}
\affiliation{Department of Physics, ETH Z\"urich, CH-8093 Z\"urich, Switzerland} 
\author{T. Ihn}
\affiliation{Department of Physics, ETH Z\"urich, CH-8093 Z\"urich, Switzerland}

\begin{abstract}
We investigate spin states of few electrons in a double quantum dot by coupling them weakly to a magnetic field resilient NbTiN microwave resonator. We observe a reduced resonator transmission if resonator photons and spin singlet states interact. This response vanishes in a magnetic field once the quantum dot ground state changes from a spin singlet into a spin triplet state. Based on this observation, we map the two-electron singlet-triplet crossover by resonant spectroscopy. By measuring the resonator only, we observe Pauli spin blockade known from transport experiments at finite source-drain bias and detect an unconventional spin blockade triggered by the absorption of resonator photons.
\end{abstract}

\maketitle

Single electrons confined in quantum dots reveal quantum effects at a fundamental level \cite{Kastner1992}. The electron wave function can be engineered to investigate phenomena due to the Pauli exclusion principle, such as exchange interaction \cite{Tarucha1996} or spin blockade \cite{Ono2002, Johnson2005}. To deduce information about such phenomena, one can couple the system to electron reservoirs and measure the resulting current \cite{Nowack2007} or utilize a charge sensor to infer information about the charge state of the system \cite{Elzerman2003}. An alternative approach is to probe the photon transmission through a microwave cavity coupling weakly to the electronic states in the quantum dots \cite{Childress2004, Burkard2016}. This approach has been used to study charge related phenomena \cite{Frey2012,Petersson2012,Delbecq2011,Toida2013,Deng2015} and valley physics \cite{Mi2017} in quantum dots. 

Here, we use such a setup with a magnetic-field-resilient resonator to study singlet-triplet spin physics in a double quantum dot (DQD). In our experiment the resonator photons interact only with the singlet states and are therefore sensitive to the singlet ground state occupation. In the first experiment, we tune the DQD into the two-electron regime where spin-singlet and -triplet states are relevant. The singlet-states form a charge qubit \cite{Hayashi2003} based on the $(1,1)$ and $(0,2)$ charge configurations. By applying an external magnetic field we enhance the resonator transmission at the bare resonator frequency indicating the transition from the singlet to the triplet ground state in the DQD. In previous experiments \cite{Schroer2012,House2015}, this transition was observed in the dispersive regime only. Here, we tune the qubit energy above or below the resonator energy and perform both resonant and dispersive spectroscopy. In contrast to Ref.~\onlinecite{Petta2005}, we map the two-electron singlet-triplet crossover at finite magnetic field with resonant spectroscopy without the need of pulsed gate operations. 

In a second experiment, we apply a finite source-drain bias and detect the spin blockade previously observed in transport experiments \cite{Ono2002, Johnson2005} by measuring the resonator only. We also detect an unconventional spin blockade that involves the absorption of resonator photons. We use a rate equation model to estimate from the finite bias data information about the spin-flip rate and the tunneling rates to the reservoirs. Knowledge of these rates contributes to fine-tuning of qubits in future quantum information processing architectures without the need of transport measurements. 

The experiments are performed with a device whose central part is shown in Fig.~1(a). We define a DQD by depleting the two dimensional electron gas (2DEG) in a GaAs/AlGaAs heterostructure locally with Au gate electrodes.
\begin{figure}[b]
\vspace{-.5cm}
\includegraphics[bb=0 0 244 120, width=\linewidth]{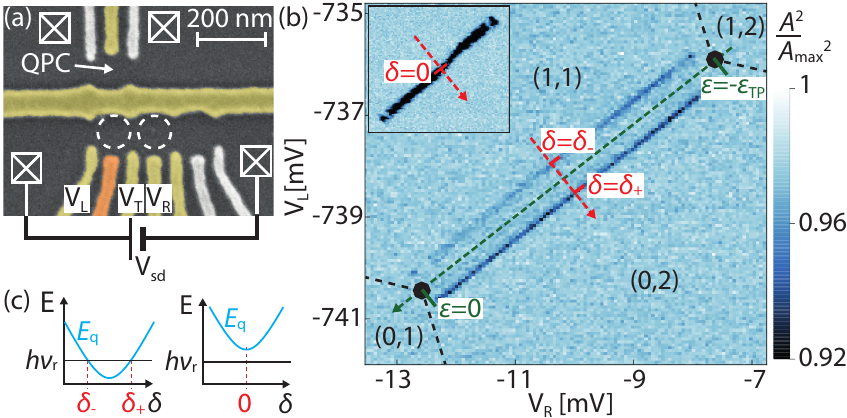}
\caption{\label{Fig1}(a) False-color scanning electron micrograph of the DQD region of the sample. The gate fingers marked in color are used to form a DQD (circles). One of the gate lines (orange) extends to the microwave resonator. Ohmic contacts to the 2DEG are indicated with squares. A voltage $V_{\mathrm{sd}}$ can be applied over the DQD. (b) Normalized resonator transmission $(A/A_\mathrm{max})^2$ on resonance as a function of $V_\mathrm{L}$ and $V_\mathrm{R}$ at $B=0$ and $V_\mathrm{sd}=0$. The inter-dot tunnel coupling is $t/h=3.4\,\mathrm{GHz}$ ($2t<h\nu_{\mathrm{r}}$) for the main figure and $t/h=4.5\,\mathrm{GHz}$ ($2t>h\nu_{\mathrm{r}}$) for the inset. Changing voltages along the dashed green line or the dashed red line independently tunes $\varepsilon$ or $\delta$, respectively. The two triple points are each marked with a black dot. (c) Bare charge qubit energy $E_\mathrm{q}$ (blue) and resonator energy $h\nu_\mathrm{r}$ (black) as a function of $\delta$ for $2t<h\nu_{\mathrm{r}}$ (left) and $2t>h\nu_{\mathrm{r}}$ (right).} 
\end{figure}
A source-drain bias $V_\mathrm{sd}$ can be applied to the DQD. The voltages $V_\mathrm{L}$ and $V_\mathrm{R}$ control the charge occupation of the DQD. The DQD charge state is determined by a quantum point pointact (QPC). One of the gate electrodes (orange in Fig.~1(a)) is electrically connected to one end of a $\lambda/2$ coplanar-waveguide resonator \cite{Frey2012} with a resonance frequency $\nu_\mathrm{r}=8.33\,\mathrm{GHz}$ and a linewidth $\kappa/2\pi=101\,\mathrm{MHz}$. The resonator is fabricated from a $15\,\mathrm{nm}$ thin film of NbTiN, which makes it resilient to parallel external magnetic fields of up to $B=2\,\mathrm{T}$ \cite{Samkharadze2016}. 

We tune the DQD into a regime where the relevant charge states are $(0,1)$, $(0,2)$, $(1,1)$ and $(1,2)$ with $(N_\mathrm{L},N_\mathrm{R})$ denoting $N_\mathrm{L}$ ($N_\mathrm{R}$) electrons in the left (right) quantum dot. 
The $(0,2)$ and $(1,1)$ charge states with singlet spin configuration form a charge qubit \cite{Hayashi2003} as they hybridize due to a tunnel coupling $t$ between the quantum dots. The qubit energy $E_\mathrm{q}=\sqrt{\delta^2+(2t)^2}$ can be tuned electrostatically: the voltages $V_\mathrm{L}$ and $V_\mathrm{R}$ control the bare energy detuning $\delta\equiv E(0,2)-E(1,1)$ of the $(0,2)$ and $(1,1)$ charge states, and the voltage $V_\mathrm{T}$ in Fig.~1(a) determines the tunnel coupling. For $2t\leq h\nu_\mathrm{r}$, the qubit and resonator energies intersect at the resonant detuning values $\delta_\mathrm{\pm}=\pm\sqrt{(h\nu_\mathrm{r})^2-(2t)^2}$ (see left panel in Fig.~1(c)). By probing the resonator transmission at the bare resonance frequency $\nu_\mathrm{r}$ as a function of $V_\mathrm{L}$ and $V_\mathrm{R}$ at $B=0$, we observe two lines with reduced transmission at $\delta_\pm$ in the main panel in Fig.~1(b). They are due to electric dipole interaction of the resonator electric field with the charge qubit in the DQD, causing a shift of the resonator resonance frequency \cite{Childress2004}. We estimate a coupling strength of $g_\mathrm{c}/2\pi\simeq28\,\mathrm{MHz}$ and a qubit decoherence rate of $\gamma_2/2\pi\simeq357\,\mathrm{MHz}$ \cite{Supplement} by using an input-output theory model \cite{Burkard2016}. Since $g_\mathrm{c}\ll\gamma_2,\kappa$, the resonator acts as a weakly coupled probe that does not influence the DQD states coherently \cite{Koski2018}.
For $2t > h\nu_\mathrm{r}$, the resonator interacts dispersively with the qubit (see right panel in Fig.~1(c)), evident as a single line with reduced transmission at $\delta = 0$ in the inset of Fig.~1(b). In the following we refer to the reduced transmission as the \enquote{resonator response}.

From Fig.~1(b) we identify the two triple points at $\varepsilon=\delta=0$, where $E(0,2)=E(1,1)=E(0,1)$, and $\varepsilon=-\varepsilon_{\mathrm{TP}}$, $\delta=0$, where $E(0,2)=E(1,1)=E(1,2)$. The system is parametrized with $\varepsilon\equiv 0.5(E(1,1)+E(0,2))-E(0,1)$. $\varepsilon_{\mathrm{TP}}\simeq510\,\mathrm{\mu eV}\,(123\,\mathrm{GHz})$ quantifies a combination of inter-dot capacitive and tunnel coupling. For $-\varepsilon_\mathrm{TP}<\varepsilon < 0 $ the charge qubit states are isolated from the reservoirs and a resonator response is visible in Fig.~1(b).

In our first experiment we probe the resonator transmission at $\nu_\mathrm{p}\simeq\nu_\mathrm{r}$ as a function of $\delta$ and $B$ for $\varepsilon\simeq-\varepsilon_{TP}/2$ and $V_\mathrm{SD}=0$. Fig.~2(a) shows the experimental result for $t/h= 4.2\,\mathrm{GHz}$ ($2t>h\nu_\mathrm{r}$), where the qubit-resonator interaction is dispersive. Like in the inset of Fig.~1(b) we observe a single dip in the resonator transmission at $\delta=0$ for $B=0$. 
\begin{figure}[h!]
\includegraphics[bb=0 0 244 319, width=\linewidth]{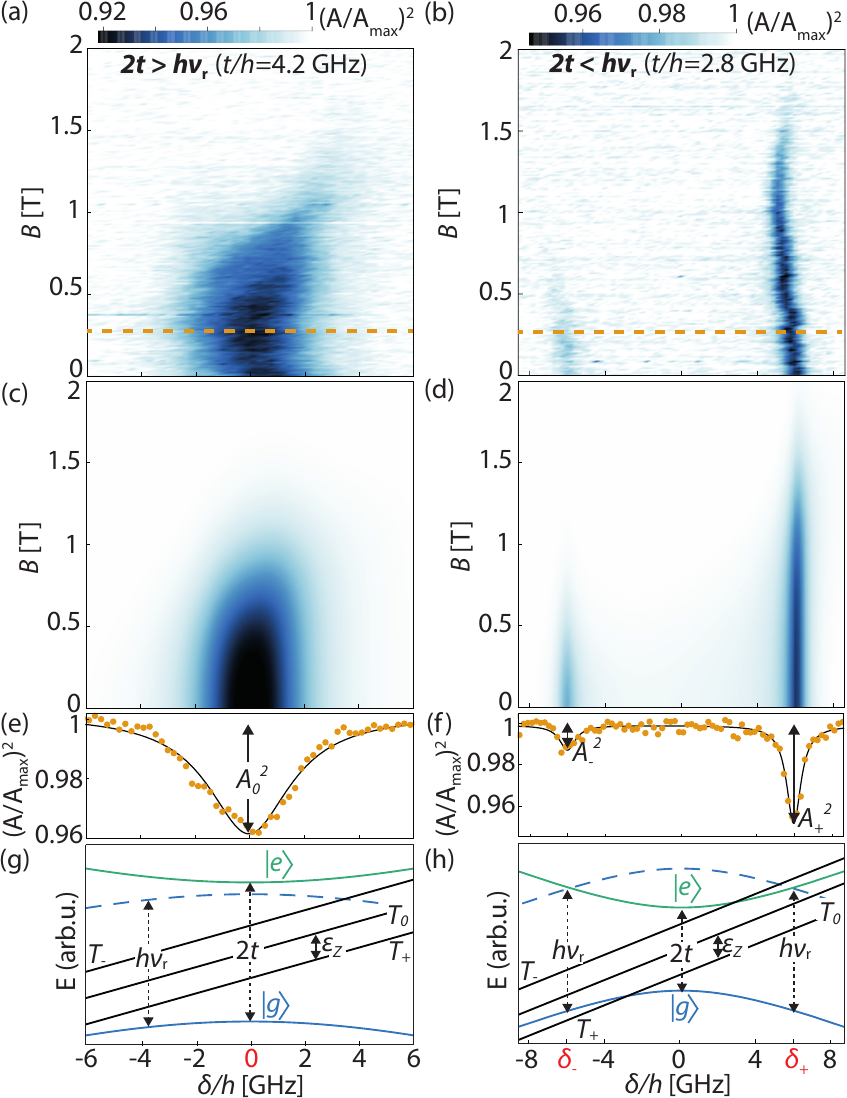}
\caption{\label{Fig2}The left (right) column is for $2t>h\nu_\mathrm{r}$ ($2t<h\nu_\mathrm{r}$). (a)-(b) Normalized resonator transmission at $\nu_\mathrm{p}\simeq \nu_\mathrm{r}$ as a function of $\delta$ and $B$. Here, $A_\text{max}$ is the bare resonance amplitude for a given $B$. (c)-(d) Theory calculation for the experimental parameters in (a)-(b). (e) Cut of the data in (a) at $B=300\,\mathrm{mT}$. The amplitude $A_0^2$ is extracted from a Lorentzian fit (solid line). (f) Cut of the data in (b) at $B=300\,\mathrm{mT}$ with a fit to two Lorentzian lineshapes (solid line) with amplitudes $A_{\pm}^2$ at the qubit-resonator resonance positions $\delta\approx\delta_\mathrm{\pm}$. (g)-(h) Schematic singlet-triplet energy spectrum as a function of detuning $\delta$ for $B=300\,\mathrm{mT}$. The spin singlet qubit ground and excited states are labeled as $\ket{g}$ (blue) and $\ket{e}$ (green), respectively. The spin triplet states are marked in black. The dashed line is offset from the qubit ground state by the resonator energy.} 
\vspace{-.5cm}
\end{figure}
The dip vanishes along a slanted line at $B\simeq1\,\mathrm{T}$ in  Fig.~2(a). For $2t<h\nu_\mathrm{r}$ shown in Fig.~2(b), the two transmission dips at $B=0$ (see Fig.~1(b)) vanish at different magnetic fields. The experimental observations in Figs.~2(a-b) are in good agreement with input-output theory calculations, shown in Figs.~2(c-d) \cite{Supplement}. 

We can qualitatively explain the observation in Figs.~2(a-b) by considering the spin character of the two electron DQD states. For $(0,2)$ only the spin singlet $(0,2)S$ is relevant since the $(0,2)T$ is about $1\,\mathrm{meV}\,(240\,\mathrm{GHz})$ higher in energy. For the $(1,1)$ charge configuration, one singlet state $(1,1)S$ and three triplet states $(1,1)T_0$, $(1,1)T_+$ and $(1,1)T_-$ are relevant.  
The resulting singlet-triplet energy spectrum as a function of $\delta$ is shown in Fig.~2(g) for $2t>h\nu_r$ and in Fig.~2(h) for $2t<h\nu_r$. Note that we neglect singlet-triplet mixing due to spin-orbit coupling or hyperfine interaction \cite{Stepanenko2012} because these mechanisms are weak compared to all other energy scales in our system. The resonator weakly probes the ground state of the system at the detuning values $\delta_\pm$ and $\delta=0$ for $2t<h\nu_\mathrm{r}$ and $2t\geq h\nu_\mathrm{r}$, respectively. For $B=0$, the ground state of the system is the qubit ground state $\ket{g}$. When increasing the magnetic field, $(1,1)T_+$ is lowered in energy with respect to $\ket{g}$ by the Zeeman energy $\varepsilon_\text{Z}$ and eventually becomes the ground state at $\delta=0$ and $\delta_\pm$. As $(1,1)T_+$ has equal charge distribution and therefore no electrical dipole moment, the resonator response at $\delta=0$ in Fig.~2(a) and at $\delta_\pm$ in Fig.~2(b) vanishes. In the following we call this phenomenon \enquote{spin blockade of the resonator response} in analogy to the spin blockade phenomenon in electron transport \cite{Ono2002}.

Figures~2(e-f) show cuts of the data in Figs.~2(a-b) at $B=300\,\mathrm{mT}$. The transmission as a function of $\delta$ has a Lorentzian lineshape with either a single dip of amplitude $A_\mathrm{0}^2$ at $\delta=0$ for $2t>h\nu_\mathrm{r}$ in Fig.~2(e) or two dips of amplitudes $A_\mathrm{\pm}^2$ at $\delta_\mathrm{\pm}$ for $2t<h\nu_\mathrm{r}$ in Fig.~2(f). To quantitatively analyze the observations in Figs.~2(a-b), we extract the magnetic field dependence of the amplitudes $A_\mathrm{0}^2$ and $A_\mathrm{\pm}^2$. The result is shown for $2t<h\nu_\mathrm{r}$ ($2t>h\nu_\mathrm{r}$) in the top (bottom) panel of Fig.~3(a).
\begin{figure}[b]
\includegraphics[bb=0 0 244 106, width=\linewidth]{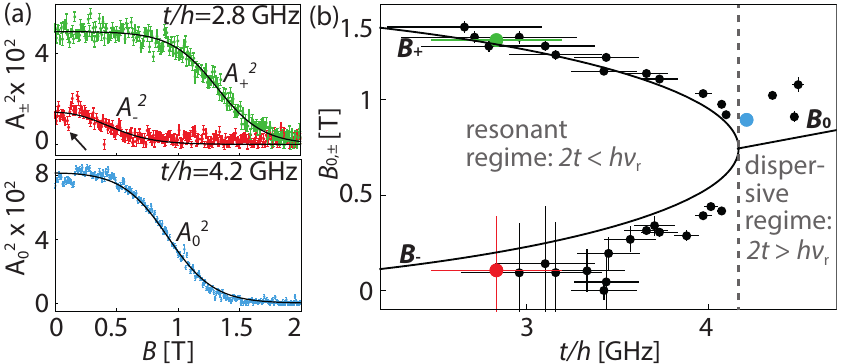}
\caption{\label{Fig3} (a) Magnetic field dependence of $A_{0}^2$ (blue), $A_{+}^2$ (green) and $A_{-}^2$ (red) with theory curve (solid line). The standard error of the fits is indicated. (b) Tunnel coupling dependence of $\ket{g}-(1,1)T_+$ intersection fields $B_0$ and $B_\pm$ compared to theory (solid line). The datapoints extracted from the fits in (a) are marked in color. The error bars in $t$ account for the error of the $\delta$ lever arm. The error bars for $B_\mathrm{0,+}$ are the standard errors of fits as in (a). For $B_-$, we show maximum error estimates from repeated measurements.}
\end{figure}
Since the resonator acts as a weak probe for singlet state transitions at $\delta=0$ or $\delta=\delta_\pm$, we find with Fermi's Golden rule that its transmission is proportional to the ground state occupation probability $p_{\ket{g}}$ \cite{Supplement}. For thermal occupation of the DQD states $p_{\ket{g}}$ is given by 
\begin{equation*}
p_{\ket{g}}(B_\delta)=1/\left(1+e^{\frac{g\mu_B B_\delta}{k_\mathrm{B} T}}+e^{\frac{g\mu_B (B-B_\delta)}{k_\mathrm{B} T}}+e^{\frac{g\mu_B (B+B_\delta)}{k_\mathrm{B} T}}\right),
\end{equation*}
where $B_\delta$ is the $\ket{g}-(1,1)T_+$ intersection field at detuning $\delta$, i.e.~for $B>B_\delta$, $(1,1)T_+$ is lower in energy than $\ket{g}$ at detuning $\delta$ (c.f.~Fig.~2(g)-(h)). From a fit of the function $C\cdot p_\mathrm{\ket{g}}$ to the magnetic field dependent amplitudes in Fig.~3(a) we extract the $\ket{g}-(1,1)T_+$ intersection fields $B_+$ at $\delta_+$ and $B_-$ at $\delta_-$ for resonant interaction (top panel) and $B_0$ at $\delta=0$ for dispersive qubit-resonator interaction (bottom panel). The proportionality constant $C$ as well as the intersection fields $B_0$ and $B_\pm$ are free parameters. Fixed parameters are the electron temperature $T_{\mathrm{e}}\simeq60\,\mathrm{mK}\,(1.3\,\mathrm{GHz})$ and the tunnel coupling $t$ obtained from the input-output theory fit at $B=0$ discussed above in the context of Fig.~1(b). We use a g-factor $g=-0.4$, which is consistent with previous work on spins in GaAs \cite{Nowack2007}.

A summary of the analysis for multiple $t$ is shown in Fig.~3(b). We obtain three branches for the three $\ket{g}-(1,1)T_+$ intersection fields $B_+$, $B_-$ and $B_0$. The values of $g\mu_B B_{\pm}$ are a direct spectroscopic measurement of the $\ket{g}-(1,1)T_+$ intersection point. The theory curve in Fig.~3(b) is calculated from the singlet-triplet energy spectrum (c.f.~Figs.~2(g-h)). There is a good agreement between this model and the experimental data over a large range of tunnel couplings. For some tunnel coupling strengths we observe a non-monotonic decrease of $A_-^2$ with magnetic field (see dip in $A_-^2$ data in Fig.~3(a) marked with an arrow), which is potentially due to a change in qubit coherence as a result of $\ket{g}-(1,1)T_+$ mixing \cite{Supplement}.

Next, we perform our second experiment by applying a finite source-drain bias across the DQD. We configure $t/h=3.7\,\mathrm{GHz}$ ($2t<h\nu_\mathrm{r}$) which gives $\delta_\pm/h=\pm 3.8\,\mathrm{GHz}$ and measure the resonant resonator transmission shown in Figs.~4(a-b) as a function of $\varepsilon$ and $\delta$ for source-drain biases $V_\text{sd} =\pm300\,\mathrm{\mu V}$ at $B=800\,\mathrm{mT}$. In this configuration, $\varepsilon_\mathrm{TP}$ and $e V_\mathrm{SD}$ are the dominant energies as $\varepsilon_\mathrm{TP}=123\,\mathrm{GHz}>e V_\mathrm{SD}=72.5\,\mathrm{GHz}\gg h\nu_r\,=8.3\,\mathrm{GHz}>2t=7.4\,\mathrm{GHz}>\varepsilon_\text{Z}=4.5\,\mathrm{GHz}>k_\mathrm{B} T=1.3\,\mathrm{GHz}$. The bias window is relevant in Fig.~4(a-b) where transport through the DQD that involves the $(0,1)$ (region A) and $(1,2)$ (region C) charge states occurs. At $\delta=0$, the extent of the bias window in $\varepsilon$ is $e V_\mathrm{sd}$. Note, that dc current through the DQD is below the limit that can be detected with our setup ($\simeq 1\,\mathrm{pA}$). A resonator response can be seen in Figs.~4(a-b) at $\delta=\delta_\pm$ like in the thermal case (c.f.~Fig.~1(b)). However, the response is suppressed in certain intervals of $\varepsilon$. While most properties of the signal are immediately evident from the corresponding energy diagrams, we highlight the regimes which are related to spin blockade in Figs.~4(a-b). The tunneling processes leading to the observations in these regimes are schematically illustrated in Figs.~4(c-e). All other regimes are discussed in the supplement. We omit the hybridization of the $(0,2)$ and $(1,1)$ charge states as well as the Zeeman splitting of $(0,1)$ and $(1,2)$ for the qualitative discussion below, but consider them in the simulations. 

We first discuss region B in Fig.~4(a-b). There, the DQD has a two-electron ground state and the bias is irrelevant. Hence, the zero bias situation discussed above applies: we observe a resonator response at $\delta_+$ and spin blockade of the resonator response at $\delta_-$ (c.f.~Fig.~2(b) for $B\geq 0.5\,\mathrm{T}$). The spin blockade of the resonator response is lifted for positive bias in the region outlined in green in Fig.~4(a) by increasing $\varepsilon$ such that $(0,1)$ is within the bias window (region A). At $\varepsilon\simeq-|e V_\mathrm{sd}/2|+h\nu_\mathrm{r}/2$ (green star in Fig.~4(a)), the electrochemical potential of $(1,1)T_+$ is aligned with $\mu_{\mathrm{d}}$ as indicated in Fig.~4(c). Hence, the tunneling sequence $(1,1)T_+\rightarrow(0,1)\rightarrow (0,2)S$ is possible (see 1.-2.~in Fig.~4(c)), which leaves the system in $(0,2)S$ that can make a transition to $(1,1)S$ by photon emission into the resonator or by phonon emission into the substrate (3.~in Fig.~4(c)). When increasing $\varepsilon$, the DQD electrochemical potentials indicated in Fig.~4(c) rise. The above tunneling sequence is possible until $\varepsilon\simeq|e V_\mathrm{sd}/2|-h\nu_\mathrm{r}/2$, where the electrochemical potential of $(0,2)S$ is on resonance with $\mu_\mathrm{S}$ (upper edge of green region in Fig.~4(a)). For larger $\varepsilon$, $(0,1)\rightarrow(0,2)S$ is suppressed and the $(0,1)$ state becomes the ground state of the system, which does not interact with the resonator.

For symmetric tunneling rates $\Gamma_\mathrm{L}$ to the left and $\Gamma_\mathrm{R}$ to the right reservoir, both region A and region C would have the same resonator response \cite{Supplement}, as transport cycles of electrons that involve $(0,1)$ (region A) are symmetric to transport cycles of holes that involve $(1,2)$ (region C). However, we have chosen asymmetric rates $\Gamma_\mathrm{R}\gg\Gamma_\mathrm{L}$ for our measurements such that $\Gamma_\mathrm{L}$ is comparable to the spin-flip rate $\Gamma_\mathrm{s}$. This allows us to quantify the ratio $\Gamma_\mathrm{s}$/$\Gamma_\mathrm{L}$ as we will show below. For positive bias, where electrons enter the DQD from the right lead (c.f.~Fig.~4(c)), the asymmetry leads to a dominant population of the $(1,2)$ charge state in region C, which does not interact with the resonator.

Next we discuss region C at negative bias, enclosed by an orange rectangle in Fig.~4(b), where we observe the spin blockade usually considered in transport experiments. In this region, transport through the DQD that involves $(1,2)$ is possible. At $\varepsilon\simeq -\varepsilon_{\mathrm{TP}}+|e V_\mathrm{sd}/2|$ (orange star), the transition $(0,2)S\rightarrow(1,2)$ is resonant with $\mu_\mathrm{d}$, i.e.~$\mu_\mathrm{d}=\mu((1,2)_{02S})$. Hence, an electron can enter the DQD from the left lead (1.~in Fig.~4(d)). Since $\mu((1,2)_{11T+})\geq\mu_{\mathrm{s}}$, $(1,2)$ can make a transition to $(1,1)T_+$ by tunneling out to the right reservoir (2.~in Fig.~4(d)). Consequently, the system is spin blocked in $(1,1)T_+$, as it can only make a transition to $(0,2)S$ by a spin-flip (3.~in Fig.~4(d)). Since $(1,1)T_+$ is higher in energy than $(0,2)S$, this corresponds to the transport spin blockade, which we observe as a reduced resonator response. Note that the response is nonzero as the left reservoir tunneling rate $\Gamma_\mathrm{L}$ is comparable to the spin-flip (singlet-triplet relaxation) rate. Therefore, the spin-flip processes $(1,1)T_+\rightarrow(0,2)S$ (3.~in Fig.~4(d)) and the tunneling cycle $(0,2)S\rightarrow(1,2)\rightarrow (1,1)T_+$ (1.-2.~in Fig.~4(d)) have similar event frequencies and $(1,1)T_+$ and $(0,2)S$ are occupied with similar probabilities. 
Note, that for symmetric reservoir barriers, the transport spin blockade would also be visible in region A, where $(0,1)$ is within the bias window \cite{Supplement}. 
\begin{figure}[t]
\includegraphics[bb=0 0 242 200, width=\linewidth]{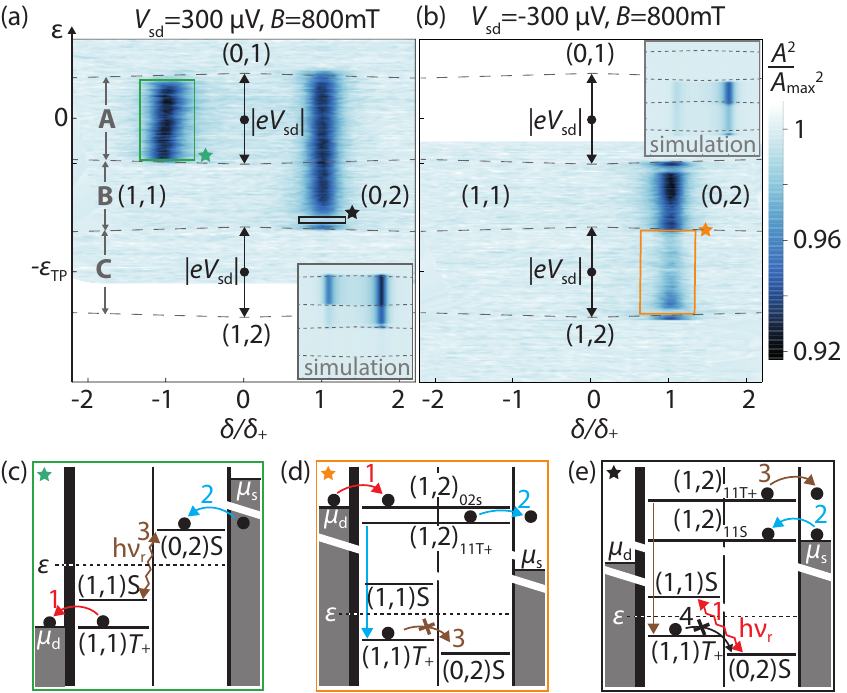}
\caption{\label{Fig4} (a-b) Normalized resonator transmission at $\nu_\mathrm{p}\simeq\nu_\mathrm{r}$ as a function of $\delta/\delta_+$ for  $B=800\,\mathrm{mT}$, $t/h=3.7\,\mathrm{GHz}$, $V_\mathrm{sd}=300\,\mathrm{\mu V}$ in (a) and $V_\mathrm{sd}=-300\,\mathrm{\mu V}$ in (b). The zero bias triple points are indicated as black dots. The inset shows theory for $\Gamma_\mathrm{R}=100\,\mathrm{MHz}$, $\Gamma_\mathrm{L}=1\,\mathrm{MHz}$, $\Gamma_\mathrm{s}=1.3\,\mathrm{MHz}$, $t/h=3.7\,\mathrm{GHz}$, and $\varepsilon_\mathrm{TP}=510\,\mathrm{\mu eV}$. (c)-(e) Energy level diagrams indicating the source ($\mu_\mathrm{s}$), drain ($\mu_\mathrm{d}$) and DQD state electrochemical potentials at $\varepsilon$ indicated by the corresponding star in (a-b). The electrochemical potentials for the DQD states are indicated with respect to $(0,1)$ for two electron states and with respect to the state $x$ for the three-electron states $(1,2)_x$. A cross marks transitions that are not possible without a spin-flip.
}
\vspace{-.5cm}
\end{figure}

In transport experiments, the spin blockade is absent if the sign of the bias is reversed. Here, we observe transport spin blockade also for positive bias at $\delta_+$ in the region indicated in black in Fig.~4(a). At $\varepsilon\simeq -\varepsilon_{\mathrm{TP}}+|e V_\mathrm{sd}/2|+h\nu_\mathrm{r}/2$ shown in Fig.~4(e), an electron can enter the DQD from the right lead and occupy $(1,2)$ because $\mu((1,2)_{1,1S})=\mu_\mathrm{s}$ (2.~in Fig.~4(e)). Compared to the transport spin blockade discussed above, this tunneling process is triggered by absorption of a resonator photon which excites the $(0,2)S$ to $(1,1)S$ transition (1.~in Fig.~4(e)). In a subsequent step, an electron can tunnel out to the right lead and leave the system in $(1,1)T_+$ (3.~in Fig.~4(e)). This realizes an unconventional spin blockade, which is triggered by an excited state transition via resonator photon absorption. The spin blockade can be lifted by a spin-flip to $(0,2)S$ (4.~Fig.~4(e)). By decreasing $\varepsilon$, tunneling into $(1,1)T_+$ is possible until $\mu((1,2)_{11T+})=\mu_{\mathrm{s}}$ (lower boundary of black rectangle in Fig.~4(a)).

We perform a classical rate equation simulation and determine the resonator transmission with input-output theory \cite{Supplement}. The simulation results shown in the insets of Figs.~4(a-b) are in good agreement with the experimental observations for a reservoir coupling asymmetry of $\Gamma_\mathrm{R}/\Gamma_\mathrm{L}\geq 100$ and a spin-flip to left lead tunneling rate ratio of $\Gamma_\mathrm{s}/\Gamma_\mathrm{L}\simeq 1$. The asymmetry was chosen such that a resonator response is absent in the simulations in region C for positive bias. The ratio $\Gamma_\mathrm{s}/\Gamma_\mathrm{L}$ is determined by the magnitude of the resonator response in the transport spin blockade situation for negative bias (orange region in Fig.~4(b)) \cite{Supplement}. Hence our experiments allow us to quantitatively estimate the tunneling rates as typical spin-flip rates for our experimental parameters are on the order of MHz \cite{Danon2013}.

In conclusion, we have studied spin physics in a few electron DQD using a weakly coupled microwave resonator as a probe in the resonant and dispersive regime. We observed spin blockade of the resonator response and could map out the two-electron singlet-triplet crossover in continuous wave experiments without the need for pulsed gate operations. In finite bias measurements we observed the conventional transport spin blockade as well as an unconventional spin blockade that is triggered by resonator photons. Signatures in the finite bias data gave direct access to relevant qubit parameters that are not easily accessible in transport experiments in the few electron regime: the symmetry of the reservoir tunneling barriers and their ratio to the spin-flip rate. The experiments presented in this work can be implemented in any other material system to investigate spin-dependent material properties like spin-orbit coupling, spin relaxation rates or the g-factor. Future work could also focus on studying the spin physics of more complex quantum states in multi quantum dot systems such as two dimensional arrays of quantum dots \cite{Mortemousque2018}. 

\begin{acknowledgments}
We would like to acknowledge fruitful discussions with Guido Burkard and Daniel Loss. This work was supported by the Swiss National Science Foundation through the National Center of Competence in Research (NCCR) Quantum Science and Technology.
\end{acknowledgments}


\begin{thebibliography}{24}%
\makeatletter
\providecommand \@ifxundefined [1]{%
 \@ifx{#1\undefined}
}%
\providecommand \@ifnum [1]{%
 \ifnum #1\expandafter \@firstoftwo
 \else \expandafter \@secondoftwo
 \fi
}%
\providecommand \@ifx [1]{%
 \ifx #1\expandafter \@firstoftwo
 \else \expandafter \@secondoftwo
 \fi
}%
\providecommand \natexlab [1]{#1}%
\providecommand \enquote  [1]{``#1''}%
\providecommand \bibnamefont  [1]{#1}%
\providecommand \bibfnamefont [1]{#1}%
\providecommand \citenamefont [1]{#1}%
\providecommand \href@noop [0]{\@secondoftwo}%
\providecommand \href [0]{\begingroup \@sanitize@url \@href}%
\providecommand \@href[1]{\@@startlink{#1}\@@href}%
\providecommand \@@href[1]{\endgroup#1\@@endlink}%
\providecommand \@sanitize@url [0]{\catcode `\\12\catcode `\$12\catcode
  `\&12\catcode `\#12\catcode `\^12\catcode `\_12\catcode `\%12\relax}%
\providecommand \@@startlink[1]{}%
\providecommand \@@endlink[0]{}%
\providecommand \url  [0]{\begingroup\@sanitize@url \@url }%
\providecommand \@url [1]{\endgroup\@href {#1}{\urlprefix }}%
\providecommand \urlprefix  [0]{URL }%
\providecommand \Eprint [0]{\href }%
\providecommand \doibase [0]{http://dx.doi.org/}%
\providecommand \selectlanguage [0]{\@gobble}%
\providecommand \bibinfo  [0]{\@secondoftwo}%
\providecommand \bibfield  [0]{\@secondoftwo}%
\providecommand \translation [1]{[#1]}%
\providecommand \BibitemOpen [0]{}%
\providecommand \bibitemStop [0]{}%
\providecommand \bibitemNoStop [0]{.\EOS\space}%
\providecommand \EOS [0]{\spacefactor3000\relax}%
\providecommand \BibitemShut  [1]{\csname bibitem#1\endcsname}%
\let\auto@bib@innerbib\@empty
\bibitem [{\citenamefont {Kastner}(1992)}]{Kastner1992}%
  \BibitemOpen
  \bibfield  {author} {\bibinfo {author} {\bibfnamefont {M.}~\bibnamefont
  {Kastner}},\ }\href {\doibase 10.1103/RevModPhys.64.849} {\bibfield
  {journal} {\bibinfo  {journal} {Rev. Mod. Phys.}\ }\textbf {\bibinfo {volume}
  {64}},\ \bibinfo {pages} {849} (\bibinfo {year} {1992})}\BibitemShut
  {NoStop}%
\bibitem [{\citenamefont {Tarucha}\ \emph {et~al.}(1996)\citenamefont
  {Tarucha}, \citenamefont {Austing}, \citenamefont {Honda}, \citenamefont
  {{van der Hage}},\ and\ \citenamefont {Kouwenhoven}}]{Tarucha1996}%
  \BibitemOpen
  \bibfield  {author} {\bibinfo {author} {\bibfnamefont {S.}~\bibnamefont
  {Tarucha}}, \bibinfo {author} {\bibfnamefont {D.G.}~\bibnamefont {Austing}},
  \bibinfo {author} {\bibfnamefont {T.}~\bibnamefont {Honda}}, \bibinfo
  {author} {\bibfnamefont {R.J.}~\bibnamefont {{van der Hage}}}, \ and\ \bibinfo {author}
  {\bibfnamefont {L.P.}~\bibnamefont {Kouwenhoven}},\ }\href {\doibase
  10.1103/PhysRevLett.77.3613} {\bibfield  {journal} {\bibinfo  {journal}
  {Physical Review Letters}\ }\textbf {\bibinfo {volume} {77}},\ \bibinfo
  {pages} {3613} (\bibinfo {year} {1996})}\BibitemShut {NoStop}%
\bibitem [{\citenamefont {Ono}(2002)}]{Ono2002}%
  \BibitemOpen
  \bibfield  {author} {\bibinfo {author} {\bibfnamefont {K.}~\bibnamefont
  {Ono}},\ }\href {\doibase 10.1126/science.1070958} {\bibfield  {journal}
  {\bibinfo  {journal} {Science}\ }\textbf {\bibinfo {volume} {297}},\ \bibinfo
  {pages} {1313} (\bibinfo {year} {2002})}\BibitemShut {NoStop}%
\bibitem [{\citenamefont {Johnson}\ \emph {et~al.}(2005)\citenamefont
  {Johnson}, \citenamefont {Petta}, \citenamefont {Marcus}, \citenamefont
  {Hanson},\ and\ \citenamefont {Gossard}}]{Johnson2005}%
  \BibitemOpen
  \bibfield  {author} {\bibinfo {author} {\bibfnamefont {A.~C.}\ \bibnamefont
  {Johnson}}, \bibinfo {author} {\bibfnamefont {J.~R.}\ \bibnamefont {Petta}},
  \bibinfo {author} {\bibfnamefont {C.~M.}\ \bibnamefont {Marcus}}, \bibinfo
  {author} {\bibfnamefont {M.~P.}\ \bibnamefont {Hanson}}, \ and\ \bibinfo
  {author} {\bibfnamefont {A.~C.}\ \bibnamefont {Gossard}},\ }\href {\doibase
  10.1103/PhysRevB.72.165308} {\bibfield  {journal} {\bibinfo  {journal}
  {Physical Review B}\ }\textbf {\bibinfo {volume} {72}},\ \bibinfo {pages} {165308}
  (\bibinfo {year} {2005})}\BibitemShut {NoStop}%
\bibitem [{\citenamefont {Nowack}\ \emph {et~al.}(2007)\citenamefont {Nowack},
  \citenamefont {Koppens}, \citenamefont {Nazarov},\ and\ \citenamefont
  {Vandersypen}}]{Nowack2007}%
  \BibitemOpen
  \bibfield  {author} {\bibinfo {author} {\bibfnamefont {K.~C.}\ \bibnamefont
  {Nowack}}, \bibinfo {author} {\bibfnamefont {F.~H.~L.}\ \bibnamefont
  {Koppens}}, \bibinfo {author} {\bibfnamefont {Y.~V.}\ \bibnamefont
  {Nazarov}}, \ and\ \bibinfo {author} {\bibfnamefont {L.~M.~K.}\ \bibnamefont
  {Vandersypen}},\ }\href {\doibase 10.1126/science.1148092} {\bibfield
  {journal} {\bibinfo  {journal} {Science}\ }\textbf {\bibinfo {volume}
  {318}},\ \bibinfo {pages} {1430} (\bibinfo {year} {2007})}\BibitemShut
  {NoStop}%
\bibitem [{\citenamefont {Elzerman}\ \emph {et~al.}(2003)\citenamefont
  {Elzerman}, \citenamefont {Hanson}, \citenamefont {Greidanus}, \citenamefont
  {{van Beveren}}, \citenamefont {{De Franceschi}}, \citenamefont
  {Vandersypen}, \citenamefont {Tarucha},\ and\ \citenamefont
  {Kouwenhoven}}]{Elzerman2003}%
  \BibitemOpen
  \bibfield  {author} {\bibinfo {author} {\bibfnamefont {J.~M.}\ \bibnamefont
  {Elzerman}}, \bibinfo {author} {\bibfnamefont {R.}~\bibnamefont {Hanson}},
  \bibinfo {author} {\bibfnamefont {J.~S.}\ \bibnamefont {Greidanus}}, \bibinfo
  {author} {\bibfnamefont {{L.~H.~Willems}}\ \bibnamefont {van Beveren}},
  \bibinfo {author} {\bibfnamefont {S.}~\bibnamefont {{De Franceschi}}},
  \bibinfo {author} {\bibfnamefont {L.~M.~K.}\ \bibnamefont {Vandersypen}},
  \bibinfo {author} {\bibfnamefont {S.}~\bibnamefont {Tarucha}}, \ and\
  \bibinfo {author} {\bibfnamefont {L.~P.}\ \bibnamefont {Kouwenhoven}},\
  }\href {\doibase 10.1103/PhysRevB.67.161308} {\bibfield  {journal} {\bibinfo
  {journal} {Physical Review B}\ }\textbf {\bibinfo {volume} {67}},\ \bibinfo
  {pages} {161308} (\bibinfo {year} {2003})}\BibitemShut {NoStop}%
\bibitem [{\citenamefont {Childress}\ \emph {et~al.}(2004)\citenamefont
  {Childress}, \citenamefont {S{\o}rensen},\ and\ \citenamefont
  {Lukin}}]{Childress2004}%
  \BibitemOpen
  \bibfield  {author} {\bibinfo {author} {\bibfnamefont {L.}~\bibnamefont
  {Childress}}, \bibinfo {author} {\bibfnamefont {A.S.}~\bibnamefont
  {S{\o}rensen}}, \ and\ \bibinfo {author} {\bibfnamefont {M.~D.}\ \bibnamefont
  {Lukin}},\ }\href {\doibase 10.1103/PhysRevA.69.042302} {\bibfield  {journal}
  {\bibinfo  {journal} {Physical Review A}\ }\textbf {\bibinfo {volume} {69}},\
  \bibinfo {pages} {042302} (\bibinfo {year} {2004})}\BibitemShut {NoStop}%
\bibitem [{\citenamefont {Burkard}\ and\ \citenamefont
  {Petta}(2016)}]{Burkard2016}%
  \BibitemOpen
  \bibfield  {author} {\bibinfo {author} {\bibfnamefont {G.}~\bibnamefont
  {Burkard}}\ and\ \bibinfo {author} {\bibfnamefont {J.~R.}\ \bibnamefont
  {Petta}},\ }\href {\doibase 10.1103/PhysRevB.94.195305} {\bibfield  {journal}
  {\bibinfo  {journal} {Physical Review B}\ }\textbf {\bibinfo {volume} {94}},\
  \bibinfo {pages} {195305} (\bibinfo {year} {2016})}\BibitemShut {NoStop}%
\bibitem [{\citenamefont {Frey}\ \emph {et~al.}(2012)\citenamefont {Frey},
  \citenamefont {Leek}, \citenamefont {Beck}, \citenamefont {Blais},
  \citenamefont {Ihn}, \citenamefont {Ensslin},\ and\ \citenamefont
  {Wallraff}}]{Frey2012}%
  \BibitemOpen
  \bibfield  {author} {\bibinfo {author} {\bibfnamefont {T.}~\bibnamefont
  {Frey}}, \bibinfo {author} {\bibfnamefont {P.~J.}\ \bibnamefont {Leek}},
  \bibinfo {author} {\bibfnamefont {M.}~\bibnamefont {Beck}}, \bibinfo {author}
  {\bibfnamefont {A.}~\bibnamefont {Blais}}, \bibinfo {author} {\bibfnamefont
  {T.}~\bibnamefont {Ihn}}, \bibinfo {author} {\bibfnamefont {K.}~\bibnamefont
  {Ensslin}}, \ and\ \bibinfo {author} {\bibfnamefont {A.}~\bibnamefont
  {Wallraff}},\ }\href {\doibase 10.1103/PhysRevLett.108.046807} {\bibfield
  {journal} {\bibinfo  {journal} {Physical Review Letters}\ }\textbf {\bibinfo
  {volume} {108}},\ \bibinfo {pages} {046807} (\bibinfo {year}
  {2012})}\BibitemShut {NoStop}%
\bibitem [{\citenamefont {Petersson}\ \emph {et~al.}(2012)\citenamefont
  {Petersson}, \citenamefont {McFaul}, \citenamefont {Schroer}, \citenamefont
  {Jung}, \citenamefont {Taylor}, \citenamefont {Houck},\ and\ \citenamefont
  {Petta}}]{Petersson2012}%
  \BibitemOpen
  \bibfield  {author} {\bibinfo {author} {\bibfnamefont {K.~D.}\ \bibnamefont
  {Petersson}}, \bibinfo {author} {\bibfnamefont {L.~W.}\ \bibnamefont
  {McFaul}}, \bibinfo {author} {\bibfnamefont {M.~D.}\ \bibnamefont {Schroer}},
  \bibinfo {author} {\bibfnamefont {M.}~\bibnamefont {Jung}}, \bibinfo {author}
  {\bibfnamefont {J.~M.}\ \bibnamefont {Taylor}}, \bibinfo {author}
  {\bibfnamefont {a.~a.}\ \bibnamefont {Houck}}, \ and\ \bibinfo {author}
  {\bibfnamefont {J.~R.}\ \bibnamefont {Petta}},\ }\href {\doibase
  10.1038/nature11559} {\bibfield  {journal} {\bibinfo  {journal} {Nature}\
  }\textbf {\bibinfo {volume} {490}},\ \bibinfo {pages} {380} (\bibinfo {year}
  {2012})}\BibitemShut {NoStop}%
\bibitem [{\citenamefont {Delbecq}\ \emph {et~al.}(2011)\citenamefont
  {Delbecq}, \citenamefont {Schmitt}, \citenamefont {Parmentier}, \citenamefont
  {Roch}, \citenamefont {Viennot}, \citenamefont {F{\`{e}}ve}, \citenamefont
  {Huard}, \citenamefont {Mora}, \citenamefont {Cottet},\ and\ \citenamefont
  {Kontos}}]{Delbecq2011}%
  \BibitemOpen
  \bibfield  {author} {\bibinfo {author} {\bibfnamefont {M.~R.}\ \bibnamefont
  {Delbecq}}, \bibinfo {author} {\bibfnamefont {V.}~\bibnamefont {Schmitt}},
  \bibinfo {author} {\bibfnamefont {F.~D.}\ \bibnamefont {Parmentier}},
  \bibinfo {author} {\bibfnamefont {N.}~\bibnamefont {Roch}}, \bibinfo {author}
  {\bibfnamefont {J.~J.}\ \bibnamefont {Viennot}}, \bibinfo {author}
  {\bibfnamefont {G.}~\bibnamefont {F{\`{e}}ve}}, \bibinfo {author}
  {\bibfnamefont {B.}~\bibnamefont {Huard}}, \bibinfo {author} {\bibfnamefont
  {C.}~\bibnamefont {Mora}}, \bibinfo {author} {\bibfnamefont {A.}~\bibnamefont
  {Cottet}}, \ and\ \bibinfo {author} {\bibfnamefont {T.}~\bibnamefont
  {Kontos}},\ }\href {\doibase 10.1103/PhysRevLett.107.256804} {\bibfield
  {journal} {\bibinfo  {journal} {Physical Review Letters}\ }\textbf {\bibinfo
  {volume} {107}},\ \bibinfo {pages} {256804} (\bibinfo {year} {2011})}\BibitemShut
  {NoStop}%
\bibitem [{\citenamefont {Toida}\ \emph {et~al.}(2013)\citenamefont {Toida},
  \citenamefont {Nakajima},\ and\ \citenamefont {Komiyama}}]{Toida2013}%
  \BibitemOpen
  \bibfield  {author} {\bibinfo {author} {\bibfnamefont {H.}~\bibnamefont
  {Toida}}, \bibinfo {author} {\bibfnamefont {T.}~\bibnamefont {Nakajima}}, \
  and\ \bibinfo {author} {\bibfnamefont {S.}~\bibnamefont {Komiyama}},\ }\href
  {\doibase 10.1103/PhysRevLett.110.066802} {\bibfield  {journal} {\bibinfo
  {journal} {Physical Review Letters}\ }\textbf {\bibinfo {volume} {110}},\
  \bibinfo {pages} {066802} (\bibinfo {year} {2013})}\BibitemShut {NoStop}%
\bibitem [{\citenamefont {Deng}\ \emph {et~al.}(2015)\citenamefont {Deng},
  \citenamefont {Wei}, \citenamefont {Johansson}, \citenamefont {Zhang},
  \citenamefont {Li}, \citenamefont {Li}, \citenamefont {Cao}, \citenamefont
  {Xiao}, \citenamefont {Tu}, \citenamefont {Guo}, \citenamefont {Jiang},
  \citenamefont {Nori},\ and\ \citenamefont {Guo}}]{Deng2015}%
  \BibitemOpen
  \bibfield  {author} {\bibinfo {author} {\bibfnamefont {G.-W.}\ \bibnamefont
  {Deng}}, \bibinfo {author} {\bibfnamefont {D.}~\bibnamefont {Wei}}, \bibinfo
  {author} {\bibfnamefont {J.~R.}\ \bibnamefont {Johansson}}, \bibinfo {author}
  {\bibfnamefont {M.-L.}\ \bibnamefont {Zhang}}, \bibinfo {author}
  {\bibfnamefont {S.-X.}\ \bibnamefont {Li}}, \bibinfo {author} {\bibfnamefont
  {H.-O.}\ \bibnamefont {Li}}, \bibinfo {author} {\bibfnamefont
  {G.}~\bibnamefont {Cao}}, \bibinfo {author} {\bibfnamefont {M.}~\bibnamefont
  {Xiao}}, \bibinfo {author} {\bibfnamefont {T.}~\bibnamefont {Tu}}, \bibinfo
  {author} {\bibfnamefont {G.-C.}\ \bibnamefont {Guo}}, \bibinfo {author}
  {\bibfnamefont {H.-W.}\ \bibnamefont {Jiang}}, \bibinfo {author}
  {\bibfnamefont {F.}~\bibnamefont {Nori}}, \ and\ \bibinfo {author}
  {\bibfnamefont {G.-P.}\ \bibnamefont {Guo}},\ }\href {\doibase
  10.1103/PhysRevLett.115.126804} {\bibfield  {journal} {\bibinfo  {journal}
  {Physical Review Letters}\ }\textbf {\bibinfo {volume} {115}},\ \bibinfo
  {pages} {126804} (\bibinfo {year} {2015})}\BibitemShut {NoStop}%
\bibitem [{\citenamefont {Mi}\ \emph {et~al.}(2017)\citenamefont {Mi},
  \citenamefont {P{\'{e}}terfalvi}, \citenamefont {Burkard},\ and\
  \citenamefont {Petta}}]{Mi2017}%
  \BibitemOpen
  \bibfield  {author} {\bibinfo {author} {\bibfnamefont {X.}~\bibnamefont
  {Mi}}, \bibinfo {author} {\bibfnamefont {C.~G.}\ \bibnamefont
  {P{\'{e}}terfalvi}}, \bibinfo {author} {\bibfnamefont {G.}~\bibnamefont
  {Burkard}}, \ and\ \bibinfo {author} {\bibfnamefont {J.~R.}\ \bibnamefont
  {Petta}},\ }\href {\doibase 10.1103/PhysRevLett.119.176803} {\bibfield
  {journal} {\bibinfo  {journal} {Physical Review Letters}\ }\textbf {\bibinfo
  {volume} {119}},\ \bibinfo {pages} {176803} (\bibinfo {year} {2017})}\BibitemShut
  {NoStop}%
\bibitem [{\citenamefont {Hayashi}\ \emph {et~al.}(2003)\citenamefont
  {Hayashi}, \citenamefont {Fujisawa}, \citenamefont {Cheong}, \citenamefont
  {Jeong},\ and\ \citenamefont {Hirayama}}]{Hayashi2003}%
  \BibitemOpen
  \bibfield  {author} {\bibinfo {author} {\bibfnamefont {T.}~\bibnamefont
  {Hayashi}}, \bibinfo {author} {\bibfnamefont {T.}~\bibnamefont {Fujisawa}},
  \bibinfo {author} {\bibfnamefont {H.~D.}\ \bibnamefont {Cheong}}, \bibinfo
  {author} {\bibfnamefont {Y.~H.}\ \bibnamefont {Jeong}}, \ and\ \bibinfo
  {author} {\bibfnamefont {Y.}~\bibnamefont {Hirayama}},\ }\href {\doibase
  10.1103/PhysRevLett.91.226804} {\bibfield  {journal} {\bibinfo  {journal}
  {Physical Review Letters}\ }\textbf {\bibinfo {volume} {91}},\ \bibinfo
  {pages} {226804} (\bibinfo {year} {2003})}\BibitemShut {NoStop}%
\bibitem [{\citenamefont {Schroer}\ \emph {et~al.}(2012)\citenamefont
  {Schroer}, \citenamefont {Jung}, \citenamefont {Petersson},\ and\
  \citenamefont {Petta}}]{Schroer2012}%
  \BibitemOpen
  \bibfield  {author} {\bibinfo {author} {\bibfnamefont {M.~D.}\ \bibnamefont
  {Schroer}}, \bibinfo {author} {\bibfnamefont {M.}~\bibnamefont {Jung}},
  \bibinfo {author} {\bibfnamefont {K.~D.}\ \bibnamefont {Petersson}}, \ and\
  \bibinfo {author} {\bibfnamefont {J.~R.}\ \bibnamefont {Petta}},\ }\href
  {\doibase 10.1103/PhysRevLett.109.166804} {\bibfield  {journal} {\bibinfo
  {journal} {Physical Review Letters}\ }\textbf {\bibinfo {volume} {109}},\
  \bibinfo {pages} {166804} (\bibinfo {year} {2012})}\BibitemShut {NoStop}%
\bibitem [{\citenamefont {House}\ \emph {et~al.}(2015)\citenamefont {House},
  \citenamefont {Kobayashi}, \citenamefont {Weber}, \citenamefont {Hile},
  \citenamefont {Watson}, \citenamefont {van~der Heijden}, \citenamefont
  {Rogge},\ and\ \citenamefont {Simmons}}]{House2015}%
  \BibitemOpen
  \bibfield  {author} {\bibinfo {author} {\bibfnamefont {M.~G.}\ \bibnamefont
  {House}}, \bibinfo {author} {\bibfnamefont {T.}~\bibnamefont {Kobayashi}},
  \bibinfo {author} {\bibfnamefont {B.}~\bibnamefont {Weber}}, \bibinfo
  {author} {\bibfnamefont {S.~J.}\ \bibnamefont {Hile}}, \bibinfo {author}
  {\bibfnamefont {T.~F.}\ \bibnamefont {Watson}}, \bibinfo {author}
  {\bibfnamefont {J.}~\bibnamefont {van~der Heijden}}, \bibinfo {author}
  {\bibfnamefont {S.}~\bibnamefont {Rogge}}, \ and\ \bibinfo {author}
  {\bibfnamefont {M.~Y.}\ \bibnamefont {Simmons}},\ }\href {\doibase
  10.1038/ncomms9848} {\bibfield  {journal} {\bibinfo  {journal} {Nature
  Communications}\ }\textbf {\bibinfo {volume} {6}},\ \bibinfo {pages} {8848}
  (\bibinfo {year} {2015})}\BibitemShut {NoStop}%
\bibitem [{\citenamefont {Petta}(2005)}]{Petta2005}%
  \BibitemOpen
  \bibfield  {author} {\bibinfo {author} {\bibfnamefont {J.~R.}\ \bibnamefont
  {Petta}},\ }\href {\doibase 10.1126/science.1116955} {\bibfield  {journal}
  {\bibinfo  {journal} {Science}\ }\textbf {\bibinfo {volume} {309}},\ \bibinfo
  {pages} {2180} (\bibinfo {year} {2005})}\BibitemShut {NoStop}%
\bibitem [{\citenamefont {Samkharadze}\ \emph {et~al.}(2016)\citenamefont
  {Samkharadze}, \citenamefont {Bruno}, \citenamefont {Scarlino}, \citenamefont
  {Zheng}, \citenamefont {DiVincenzo}, \citenamefont {DiCarlo},\ and\
  \citenamefont {Vandersypen}}]{Samkharadze2016}%
  \BibitemOpen
  \bibfield  {author} {\bibinfo {author} {\bibfnamefont {N.}~\bibnamefont
  {Samkharadze}}, \bibinfo {author} {\bibfnamefont {A.}~\bibnamefont {Bruno}},
  \bibinfo {author} {\bibfnamefont {P.}~\bibnamefont {Scarlino}}, \bibinfo
  {author} {\bibfnamefont {G.}~\bibnamefont {Zheng}}, \bibinfo {author}
  {\bibfnamefont {D.~P.}\ \bibnamefont {DiVincenzo}}, \bibinfo {author}
  {\bibfnamefont {L.}~\bibnamefont {DiCarlo}}, \ and\ \bibinfo {author}
  {\bibfnamefont {L.~M.~K.}\ \bibnamefont {Vandersypen}},\ }\href {\doibase
  10.1103/PhysRevApplied.5.044004} {\bibfield  {journal} {\bibinfo  {journal}
  {Physical Review Applied}\ }\textbf {\bibinfo {volume} {5}},\ \bibinfo
  {pages} {044004} (\bibinfo {year} {2016})}\BibitemShut {NoStop}%
\bibitem [{Sup()}]{Supplement}%
  \BibitemOpen
  \href@noop {} {\bibinfo  {journal} {See Supplemental Material for the
  input-output theory and rate equation models, a discussion of singlet-triplet
  mixing and details on the finite-bias data that were not mentioned in the
  main text}\ }\BibitemShut {NoStop}%
\bibitem [{\citenamefont {Koski}\ \emph {et~al.}(2018)\citenamefont {Koski},
  \citenamefont {Landig}, \citenamefont {P{\'{a}}lyi}, \citenamefont
  {Scarlino}, \citenamefont {Reichl}, \citenamefont {Wegscheider},
  \citenamefont {Burkard}, \citenamefont {Wallraff}, \citenamefont {Ensslin},\
  and\ \citenamefont {Ihn}}]{Koski2018}%
  \BibitemOpen
\bibfield  {journal} {  }\bibfield  {author} {\bibinfo {author} {\bibfnamefont
  {J.~V.}\ \bibnamefont {Koski}}, \bibinfo {author} {\bibfnamefont {A.~J.}\
  \bibnamefont {Landig}}, \bibinfo {author} {\bibfnamefont {A.}~\bibnamefont
  {P{\'{a}}lyi}}, \bibinfo {author} {\bibfnamefont {P.}~\bibnamefont
  {Scarlino}}, \bibinfo {author} {\bibfnamefont {C.}~\bibnamefont {Reichl}},
  \bibinfo {author} {\bibfnamefont {W.}~\bibnamefont {Wegscheider}}, \bibinfo
  {author} {\bibfnamefont {G.}~\bibnamefont {Burkard}}, \bibinfo {author}
  {\bibfnamefont {A.}~\bibnamefont {Wallraff}}, \bibinfo {author}
  {\bibfnamefont {K.}~\bibnamefont {Ensslin}}, \ and\ \bibinfo {author}
  {\bibfnamefont {T.}~\bibnamefont {Ihn}},\ }\href {\doibase
  10.1103/PhysRevLett.121.043603} {\bibfield  {journal} {\bibinfo  {journal}
  {Physical Review Letters}\ }\textbf {\bibinfo {volume} {121}},\ \bibinfo
  {pages} {043603} (\bibinfo {year} {2018})}\BibitemShut {NoStop}%
\bibitem [{\citenamefont {Stepanenko}\ \emph {et~al.}(2012)\citenamefont
  {Stepanenko}, \citenamefont {Rudner}, \citenamefont {Halperin},\ and\
  \citenamefont {Loss}}]{Stepanenko2012}%
  \BibitemOpen
  \bibfield  {author} {\bibinfo {author} {\bibfnamefont {D.}~\bibnamefont
  {Stepanenko}}, \bibinfo {author} {\bibfnamefont {M.}~\bibnamefont {Rudner}},
  \bibinfo {author} {\bibfnamefont {B.~I.}\ \bibnamefont {Halperin}}, \ and\
  \bibinfo {author} {\bibfnamefont {D.}~\bibnamefont {Loss}},\ }\href {\doibase
  10.1103/PhysRevB.85.075416} {\bibfield  {journal} {\bibinfo  {journal}
  {Physical Review B}\ }\textbf {\bibinfo {volume} {85}},\ \bibinfo {pages}
  {075416} (\bibinfo {year} {2012})}\BibitemShut {NoStop}%
\bibitem [{\citenamefont {Danon}(2013)}]{Danon2013}%
  \BibitemOpen
  \bibfield  {author} {\bibinfo {author} {\bibfnamefont {J.}~\bibnamefont
  {Danon}},\ }\href {\doibase 10.1103/PhysRevB.88.075306} {\bibfield  {journal}
  {\bibinfo  {journal} {Physical Review B}\ }\textbf {\bibinfo {volume} {88}},\
  \bibinfo {pages} {075306} (\bibinfo {year} {2013})}\BibitemShut {NoStop}%
\bibitem [{\citenamefont {Mortemousque}\ \emph {et~al.}()\citenamefont
  {Mortemousque}, \citenamefont {Chanrion}, \citenamefont {Jadot},
  \citenamefont {Flentje}, \citenamefont {Ludwig}, \citenamefont {Wieck},
  \citenamefont {Urdampilleta}, \citenamefont {Bauerle},\ and\ \citenamefont
  {Meunier}}]{Mortemousque2018}%
  \BibitemOpen
  \bibfield  {author} {\bibinfo {author} {\bibfnamefont {P.-A.}\ \bibnamefont
  {Mortemousque}}, \bibinfo {author} {\bibfnamefont {E.}~\bibnamefont
  {Chanrion}}, \bibinfo {author} {\bibfnamefont {B.}~\bibnamefont {Jadot}},
  \bibinfo {author} {\bibfnamefont {H.}~\bibnamefont {Flentje}}, \bibinfo
  {author} {\bibfnamefont {A.}~\bibnamefont {Ludwig}}, \bibinfo {author}
  {\bibfnamefont {A.~D.}\ \bibnamefont {Wieck}}, \bibinfo {author}
  {\bibfnamefont {M.}~\bibnamefont {Urdampilleta}}, \bibinfo {author}
  {\bibfnamefont {C.}~\bibnamefont {Bauerle}}, \ and\ \bibinfo {author}
  {\bibfnamefont {T.}~\bibnamefont {Meunier}},\ }\href
  {http://arxiv.org/abs/1808.06180} {\bibinfo  {journal} {arXiv:1808.06180}\
  }\BibitemShut {NoStop}%
\end{thebibliography}
\end{document}